\documentclass[twocolumn,showpacs,superscriptaddress,aps,pra]{revtex4}
\usepackage{epsfig}
\usepackage{subfigure}
\usepackage{amssymb}
\usepackage{graphicx}
\usepackage{color}
\usepackage{amsfonts}
\usepackage{amscd}
\usepackage{hyperref}
\usepackage{amsmath}
\usepackage{float}
\usepackage{epstopdf}
\begin{document}
\title{Controllable coupling between a nanomechanical resonator and a coplanar-waveguide resonator via a superconducting flux qubit }
\author{Wei Xiong}
\affiliation{Department of Physics, Fudan University, Shanghai 200433, China}
\affiliation{Quantum Physics and Quantum Information Division, Beijing Computational Science Research Center, Beijing 100094, China}
\affiliation{Department of Applied Physics, Hong Kong Polytechnic University, Hung Hom,
Hong Kong, China}
\author{Da-Yu Jin}
\affiliation{Department of Physics, Fudan University, Shanghai 200433, China}
\author{Jun Jing}
\affiliation{Institute of Atomic and Molecule Physics, Jilin University, Changchun
130012, China}
\author{Chi-Hang Lam}
\altaffiliation{C.H.Lam@polyu.edu.hk}
\affiliation{Department of Applied Physics, Hong Kong Polytechnic University, Hung Hom,
Hong Kong, China}
\author{J. Q. You}
\altaffiliation{jqyou@csrc.ac.cn}
\affiliation{Quantum Physics and Quantum Information Division, Beijing Computational Science Research Center, Beijing 100094, China}
\affiliation{Synergetic Innovation Center of Quantum Information and Quantum Physics,
University of Science and Technology of China, Hefei, Anhui 230026, China}

\date{\today}

\begin{abstract}

We study a tripartite quantum system consisting of a coplanar-waveguide (CPW)
resonator and a nanomechanical resonator (NAMR) connected by a flux qubit, where the flux qubit has a large detuning from both resonators. By a unitary transformation and a second-order approximation, we obtain a strong and controllable (i.e., magnetic-field-dependent) effective coupling between the NAMR and the CPW resonator. Due to the strong coupling, vacuum Rabi splitting can be observed from the voltage-fluctuation spectrum of the CPW resonator. We further study the properties of single photon transport as inferred from the reflectance or equivalently the transmittance. We show that the reflectance and the corresponding phase shift spectra both exhibit doublet of narrow spectral features due to vacuum Rabi splitting. By tuning the external magnetic field, the reflectance and the phase shift can be varied from 0 to 1 and $-\pi$ to $\pi$, respectively. The results indicate that this hybrid quantum system can act as a quantum router.
\end{abstract}


\maketitle

\section{introduction}

Recently, nanomechanical resonators (NAMRs)~\cite{Blencowe,Schwab} have demonstrated great potential for applications in quantum measurement \cite{Braginsky} and high-precision displacement detection \cite{LaHaye,Bocko,VP}, and have attracted considerable attention. With improved techniques, high quality factors (from $10^2$ to $10^5$) and large fundamental frequencies (in the range of MHz-GHz)~\cite{Cleland,Huang,Gaidarzhy}~have been achieved. This puts the NAMR in the quantum limit in which the vibrational energy is smaller than the thermal energy, i.e., $\hbar\omega_m<K_B T$. As a consequence, NAMRs can serve as components in more sophisticated quantum systems \cite{Bennett,Rabl,Arcizet,Kolkowitz,Ouyang,zzLi,Tian1,Tian2,Aspelmeyer,Lu,Xia,Xue1} operating under a large frequency range from the microwave to the optical domain. Among these systems, a NAMR embedded in a superconducting qubit is a good candidate for exploring various quantum phenomena at the boundary between classical and quantum regimes. For example, controllable coupling \cite{Xue1}, quantum entanglement \cite{Tian3,Chen}, squeezed state \cite{Cohen,Xue2}, quantum detection \cite{Schmidt}, ground-state cooling \cite{Xia,JZhang,PZhang,JQYou} and phonon blockade \cite{YXLiu} have been studied. However, the usefulness of qubit-NAMR systems in further investigations is limited by their simple structures and the few degrees of freedom. More complex hybrid systems should thus be considered.

One possibility is to engineer a qubit-NAMR system into a coplanar-wideguide (CPW) resonator. Such a system can serve as the analog of cavity quantum electrodynamics (QED) providing a basic interaction between light and matter. It can also be used to study interesting phenomena such as quantum-classical transition \cite{LFWei}. Also, a tunable strong coupling between a microwave cavity and a NAMR mediated by a Cooper pair transistor has been studied recently \cite{MPB}. Moreover, strong coupling~\cite{Wallraff} or even ultrastrong coupling~\cite{Niemczyk} between a CPW resonator and a superconducting qubit has been achieved. This can be important for quantum information processing. Nevertheless, due to the size mismatch between typical NAMRs and CPW resonators, their direct coupling is in general weak and uncontrollable as well.

Motivated by these issues, we generalize the setups in Refs.~\cite{Wallraff,Xue1} and propose a tripartite system to realize a strong and controllable effective coupling between a NAMR and a CPW resonator via a flux qubit. The qubit has a large detuning from both the NAMR and the CPW resonator. After eliminating the degrees of freedom of the qubit using a unitary transformation and a second-order approximation, we are able to find the effective coupling between the two resonators. We further show that this effective coupling can be switched on or off and tuned from weak to very strong via controlling an external magnetic field parallel to the plane of the qubit.

We also investigate the voltage-fluctuation spectrum (VFS) of the CPW resonator in the hybrid system. We show that the VFS depends on the qubit state. The resonant peak in the spectrum splits into two in the presence of the NAMR and the qubit. The separation between the pair of peaks increases when the coupling strength between the qubit and the NAMR is enhanced by increasing the magnetic field. We further give an analysis on the reflectance of an incident photon by the CPW resonator. We show that the incident photon can have a perfect reflection in the whole frequency range in the absence of the NAMR. However, in the presence of the NAMR, the reflectance dips at resonance. Furthermore, the peak splitting occurring in the VFS  is associate with an analogous split of the dip in the reflectance. In particular, the reflectance can be tuned from 1 to 0 by varying the coupling strength between the qubit and the NAMR. Therefore, the NAMR can work as a quantum router to direct the incident photon from one channel to the other. We also discuss briefly the phase shift of a single photon traveling in the CPW resonator. Specifically, the phase-shift spectrum shows a narrow spectral feature of non-zero shift close to the resonant point. In the presence of the NAMR, this spectral feature also splits into a doublet.

The paper is organized as follows. Sec. II introduces the Hamiltonian of our hybrid quantum system. In Sec. III, we derive the effective Hamiltonian between the NAMR and the CPW resonator, and then give an estimation on this effective coupling using realistic parameters. In Sec. IV, we study the effects of the NAMR on the VFS of the CPW resonator. We further study the single photon transport in our hybrid system in Sec. V. In Sec. VI, we discuss the effect of the qubit decay on the reflection coefficient. Finally, we give a short conclusion in Sec. VII.

\begin{figure}[H]
  \centering
 \includegraphics[scale=0.7]{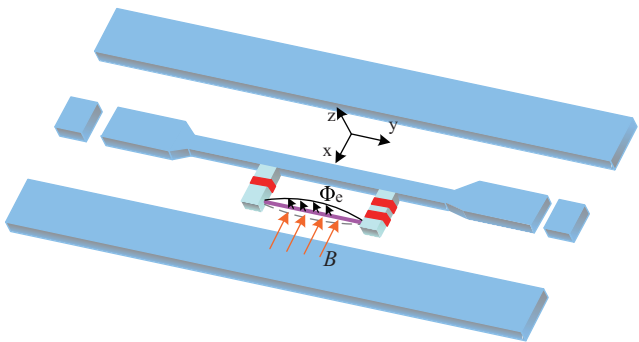}\\
 \caption{(Color online) Schematic diagram of the proposed hybrid quantum
system, which consists of a CPW resonator containing a tri-Josephson-junction flux qubit coupled to a NAMR. $\Phi _{e}$ is the magnetic flux through the qubit loop. $B$ is an external magnetic field along the $x$-direction for controlling the coupling strength between the flux qubit and the NAMR.}
\end{figure}

\section {The Hamiltonian of the hybrid system}

The proposed hybrid system 
consists of a NAMR and a CPW resonator both directly coupled to a superconducting flux qubit (see Fig.~1). Its Hamiltonian can be written as  (setting $\hbar $=1)
\begin{equation}
H=H_{c}+H_{m}+H_{q}+H_{cq}+H_{mq},  \label{system H}
\end{equation}
Here, $H_{c}=\omega _{c}c^{\dag }c$ is the Hamiltonian of the CPW resonator, where $c^\dag$ ($c$) is the creation (annihilation) operator of the resonator mode with frequency $\omega_c$. The Hamiltonian of the NAMR is $H_{m}=\omega _{m}b^{\dag }b$, with $b^\dag$ ($b$) being the creation (annihilation) operator of the vibrational mode with frequency $\omega_m$. The Hamiltonian of the superconducting flux qubit, which is composed of a superconducting loop interrupted by three Josephson junctions, is given by~\cite{Orlando}
\begin{equation}
H_{q}=\frac{1}{2}(\varepsilon \sigma _{z}+\mu \sigma _{x}),  \label{flux H}
\end{equation}%
where $\varepsilon =I_{p}(2\Phi _{e}-\Phi _{0})$, with $I_{p}$ being the persistent current, $\Phi _e$ the external magnetic flux threading the loop and $\Phi _{0}=h/2e$ the magnetic flux quantum. The operators $\sigma _{x}$, $\sigma _{y}$ and $\sigma _{z}$ are spin-$1/2$ Pauli operators. The interaction Hamiltonian $H_{cq}$ represents the coupling between the flux qubit and the CPW resonator. To achieve a strong qubit-resonator coupling, the flux qubit is placed at the antinode of the standing wave of the current on the lathy central conductor of the CPW resonator, where the magnetic field is strong. The interaction Hamiltonian $H_{cq}$ is \cite{Xiang}
\begin{equation}
H_{cq}=g_{c}(a+a^{\dag })\sigma _{z},  \label{cq H}
\end{equation}
where the coupling strength is $g_{c}=MI_{p}I_0$, with $M$ denoting the mutual inductance and $I_{0}=(\omega _c/L_c)^{1/2}$ being the zero-point current ($L_c$ is the total inductance of the resonator).
The interaction Hamiltonian $H_{mq}$ in Eq.~(\ref{system H}) represents the coupling between the flux qubit and the NAMR. As shown in Fig.~1, the circulating current in the superconducting loop experiences an external magnetic field $B$ in the $x$ direction, which is parallel to the loop of the flux qubit. Then, an Ampere's force is generated which drives the NAMR to oscillate in the $z$ direction. Thus, the flux qubit and the NAMR are coupled and the interaction Hamiltonian $H_{mq}$ is given by~\cite{Xue1}
\begin{equation}
H_{mq}=g_{m}(b+b^{\dag })\sigma _{z},  \label{NAMR-q H}
\end{equation}%
where the coupling strength is $g_{m}=BI_{p}l\delta _\text{zpm}$, with $\delta _\text{zpm}=1/\sqrt{2m\omega _{m}}$ being the zero-point motion, $m$ the effective mass and $l=\chi l_{0}$ the effective length ($l_{0}$ is the original length of the NAMR and $\chi $ is a factor depending on the oscillation mode \cite{Husain}). Therefore, the coupling strength $g_{m}$ can be directly controlled by the external magnetic field $B$.

To minimize the effect of the flux noise, we bias the flux qubit at its degeneracy point (i.e., $\varepsilon =0$). Also, we use the eigenstates of the flux qubit as the new basis states. Then, $\sigma_x\rightarrow\sigma_z$ and $\sigma_z\rightarrow\sigma_x$. Neglecting the fast oscillating terms via the rotating-wave approximation, we have
\begin{align}
H& =\omega _{c}c^{\dag }c+\omega _{m}b^{\dag }b+\frac{1}{2}\omega _{q}\sigma
_{z}  \nonumber\\
& +g_{c}(c^{\dag }\sigma _{-}+c\sigma _{+})+g_{m}(b^{\dag }\sigma
_{-}+b\sigma _{+}),  \label{total H}
\end{align}%
where $\omega _{q}=\mu$ is the transition frequency of the flux qubit at the degeneracy point. In Ref.~\cite{JMP}, a Hamiltonian similar to Eq.~(\ref{total H}) was experimentally realized using a superconducting transmon qubit simultaneously coupled to a CPW resonator and a membrane-type micromechanical resonator. Different from the dispersive regime used in our work, the hybrid system in \cite{JMP} was achieved in the resonant regime to transfer quantum information between the two resonators via the transmon qubit.

\section{The effective coupling between a NAMR and a CPW resonator}

We consider the hybrid system in the dispersive regime where the coupling strengths $g_c$ and $g_m$ are much smaller than the frequency detunings of the flux qubit from the CPW resonator ($\Delta _{c}\equiv \omega _{q}-\omega _{c}>0$) and the NAMR ($\Delta _{m}\equiv \omega_{q}-\omega _{m}>0$), respectively ({i.e.},
$g_{c}/\Delta _{c}\ll 1~\text{and }~g_{m}/\Delta _{m}\ll 1$).
Then, we can divide the Hamiltonian~(\ref{total H}) into two parts: $H=H_{0}+H_{I}$, where
\begin{equation}
H_{0}=\omega _{c}c^{\dag }c+\omega _{m}b^{\dag }b+\frac{1}{2}\omega _{q}\sigma _{z},  \label{free H}
\end{equation}%
and the perturbation part is
\begin{equation}
H_{I}=g_{c}(c^{\dag }\sigma _{-}+c\sigma _{+})+g_{m}(b^{\dag}\sigma _{-}+b\sigma _{+}).  \label{interaction H}
\end{equation}%
Here we apply a Fr\"{o}hlich-Nakajima transformation \cite{Frohlich,Nakajima,Niskanen} to the Hamiltonian~(\ref{total H}):
\begin{align}
H_\text{eff}=UHU^{\dag},\label{Effective H}
\end{align}
where
\begin{align}
U&=\exp{(-V)},\nonumber\\
V&=\eta _{c}(c^{\dag }\sigma _{-}-c\sigma _{+})+\eta _{m}(b^{\dag }\sigma_{-}-b\sigma _{+}).
\end{align}
This $V$ satisfies
\begin{equation}
H_{I}+[H_0,V]=0,  \label{condition}
\end{equation}%
which gives rise to $\eta _{c}=g_{c}/\Delta _{c}$ and $\eta_{m}=g_{m}/\Delta _{m}$. Because the coefficients $\eta _{c}$ and $\eta _{m}$ are both small, the higher-order terms can be dropped and only the second-order term $[H_{I},V]$ needs to be kept in Eq.~(\ref{Effective H}). Therefore, the effective Hamiltonian $H_\text{eff}$ is reduced to
\begin{equation}
H_{\text{eff}}\simeq H_{0}+\frac{1}{2}[H_{I},V].\label{Heff}
\end{equation}%
Using the relations $[\sigma _{+},\sigma _{-}]=\sigma _{z}$, $[\sigma _{z},\sigma
_{-}]=-2\sigma _{-}$ and $[\sigma _{z},\sigma _{+}]=2\sigma _{+}$, we can explicitly write Eq.~(\ref{Heff}) as
\begin{equation}
H_{\text{eff}}=\omega_{c}^\prime c^{\dag }c+\omega_{m}^\prime b^{\dag }b+g(c^{\dag}b+cb^{\dag }),  \label{Reduced H}
\end{equation}%
where
\begin{align}
\omega_{c}^\prime& =\omega _{c}-\frac{g_{c}^{2}}{\Delta _{c}},~~~\omega_{m}^\prime =\omega _{m}-\frac{g_{m}^{2}}{\Delta _{m}},\nonumber\\
g& =-\frac{g_{c}g_{m}}{2}(\frac{1}{\Delta _{c}}+\frac{1}{\Delta _{m}}).\label{parameters}
\end{align}%
In deriving Eq.~(\ref{Reduced H}), we have also assumed that the flux qubit is initially prepared in its ground state so that the expectation value $\langle \sigma _{z}\rangle=-1$ in the adiabatic approximation. Then we can eliminate the degrees of freedom of the flux qubit and the effective Hamiltonian~(\ref{Reduced H}) is obtained. From Eq.~(\ref{parameters}), one can see that the frequenies of the NAMR and the CPW resonator are both red-shifted due to the flux qubit, and the effective coupling strength $g$ between the NAMR and the CPW resonator depends on the coupling strengths, $g_{c}$ and $g_{m}$, and the frequency detunings, $\Delta _{c}$ and $\Delta _{m}$. Thus, the effective coupling strength $g$ is {\it controllable}.

Below we give an esitmation on this effective coupling strength $g$. For simplicity, here we assume that the NAMR and the CPW resonator are in resonance, i.e., $\omega_c=\omega_m$, and the frequency detuning between the flux qubit and the two resoantors are $\Delta_c/2\pi=\Delta_m/2\pi=1$ GHz. According to experiments \cite{Wallraff}, the coupling strength $g_c$ can be legitimately chosen as $\sim 2\pi\times100$ MHz. The coupling strength $g_m$ can also be $\sim 2\pi\times100$ MHz when using an external magnetic field $B\sim 0.125$ T and the accessible parameters \cite{Gaidarzhy,Martin,Lupascu}, $I_{p}\sim 660$ nA, $l_{0}\sim 3.9\times10^{-13}$ m, $\delta _{zpm}\sim 2.6\times 10^{-13}$ m, and $\chi \sim 0.8$. Thus, the effective coupling strength $g$ between the NAMR and the CPW resonator is $\sim 2\pi\times10$ MHz. Low decay rates were achieved in experiments~\cite{Kubo,Connell} for the CPW resoantor ($\kappa\sim 1$ MHz), the flux qubit ($\gamma_q\sim 1$ MHz), and the NAMR ($\gamma<1$ MHz). Therefore, this effective coupling $g$ can be in the strong coupling regime. Since the coupling strength $g_m$ is proportional to the external magnetic field $B$, the effective coupling strength $g$ can become larger than $2\pi\times10$ MHz when using a stronger $B$~(e.g., $g\sim 50$ MHz for $B\sim 0.5$ T). With this strong and controllable coupling $g$, quantum state transfer between the CPW resonator and the NAMR can be readily achieved. For example, when the quantum information is initially encoded as $|0\rangle_c+|1\rangle_c$ in the CPW resonator, after an evolution time $gt=(2n+1)\pi/2$ ($n=0,1,2,\ldots$), the information will be transferred to the NAMR as $(|0\rangle_c+|1\rangle_c)|0\rangle_m\rightarrow |0\rangle_c(|0\rangle_m+|1\rangle_m)$.

\begin{figure}
     \includegraphics[scale=0.55]{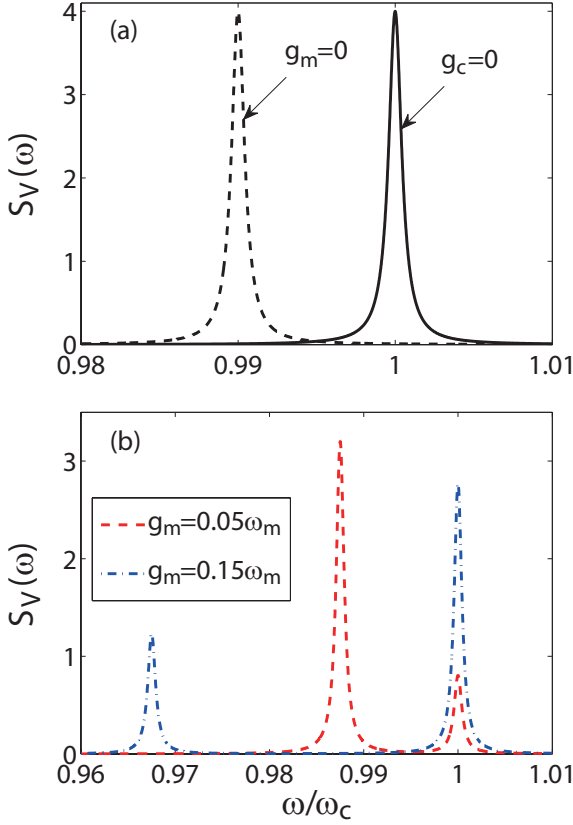}\\
  \caption{(Color online)
The voltage-fluctuation spectrum $S_V(\omega)$  of the CPW resonator as a function of the normalized frequency $\omega/\omega_c$. (a) The CPW is decoupled from the flux qubit with $g_c=0$ (solid curve) or to the flux qubit with $g_c\neq0$ but $g_m=0$ (dashed curve). (b) Both the CPW and the NAMR are coupled to the flux qubit with $\omega_m=\omega_c$, $g_c=0.1\omega_c$, and $g_m=0.05\omega_m$ (dashed curve) or $g_m=0.15\omega_m$ (dot-dashed curve). Other parameters for (a) and (b) are $\omega_q=2\omega_c$, $\kappa=10^{-3}\omega_c$, $\gamma=10^{-3}\omega_m$ and $\omega_c/2\pi=1$GHz.
}
\end{figure}

\section{Voltage-fluctuation spectrum}

In this section, we investigate the effects of the NAMR on the voltage-fluctuation spectrum (VFS) of the CPW resonator defined as \cite{LFWei}
\begin{equation}
S_{V}(\omega )=\frac{1}{2\pi }\int_{-\infty }^{+\infty }d\tau e^{i\omega
\tau }\langle V(t)V(t+\tau )\rangle _{t\rightarrow \infty },  \label{spectrum}
\end{equation}%
where $V(t)\propto c^{\dag }(t)+c(t)$ is the voltage in the CPW resonator at a given point and it can be detected
by a standard rf network analyzer \cite{Badzey}. For convenience, we substitute `$\propto$' with `=' in what follows due to the fact that the relative amplitude of the spectrum does not affect the results. To obtain the VFS, the two-time correlation $\langle V(t)V(t+\tau )\rangle$ should be calculated first. We employ quantum Langevin equations governing the dynamics of the system~\cite{Walls},
\begin{align}
\frac{dc(t)}{dt}& =-\left( i\omega_{c}^\prime+\frac{\kappa}{2}\right)
c(t)-igb(t)+\sqrt{\kappa}c_{\rm in}(t),
\nonumber \\
\frac{db(t)}{dt}& =-\left( i\omega_{m}^\prime+\frac{\gamma}{2}\right)
b(t)-igc(t)+\sqrt{\gamma}b_{\rm in}(t),\label{QLE}
\end{align}%
where $\kappa$ is the quantum state decay rate in the CPW resonator, $\gamma$ is the decay rate in the NAMR, and both $c_{\rm in}(t)$ and $b_{\rm in}(t)$ are input noise operators with zero mean values $\langle c_{\rm in}(t)\rangle =\langle b_{\rm in}(t)\rangle =0$, which are related to the CPW resonator and the NAMR, respectively. The two-time correlation functions at a low temperature are
\begin{align}
\langle c_{\rm in}(t)c_{\rm in}^{\dag }(t ^{\prime })\rangle & =\delta
(t -t^{\prime }),~~~\langle c_{\rm in}^{\dag }(t )c_{\rm in}(t^\prime
)\rangle=0,  \nonumber \\
\langle b_{\rm in}(t)b_{\rm in}^{\dag }(t ^{\prime })\rangle & =\delta
(t -t^{\prime}),~~~\langle b_{\rm in}^{\dag }(t)b_{\rm in}(t^\prime
)\rangle=0.\label{correlation}
\end{align}%
In fact, Eq.~(\ref{QLE}) can be solved more easily in the frequency domain rather than in the time domain. By performing Fourier transformations $c(t ) =\frac{1}{\sqrt{2\pi }}\int_{-\infty }^{+\infty }c(\omega)e^{i\omega
t}d\omega$ and
$b(t )=\frac{1}{\sqrt{2\pi }}\int_{-\infty }^{+\infty }b(\omega)e^{i\omega
t}d\omega$, we obtain
\begin{equation}
c(\omega )=\left[ \sqrt{\kappa}c_{\rm in}(\omega )-\frac{ig\sqrt{%
\gamma}b_{\rm in}(\omega )}{\gamma/2-i(\omega -\omega_{m}^\prime)}\right] /d\left( \omega \right) ,  \label{operator}
\end{equation}%
where
\begin{equation}
d(\omega )=\frac{\kappa}{2}-i(\omega -\omega_{c}^\prime)+\frac{g^{2}}{\gamma/2-i(\omega -\omega_{m}^\prime)}.
\end{equation}%
Then, the VFS in Eq.~(\ref{spectrum}) can be written as
\begin{equation}
S_{V}(\omega )=\frac{1}{d(\omega )}+\frac{1}{d^{\ast }(\omega )}.  \label{spectrum2}
\end{equation}
This expression means that the behavior of the VFS is completely determined by $d(\omega)$.

In Fig.~2, we plot the VFS as a function of the normalized frequency $\omega/\omega_c$. The profile of the VFS has a Lorentzian shape with a single peak when the effective coupling $g$ vanishes. However, there are two cases which give rise to $g=0$, i.e., $g_c=0$ or $g_m=0$ but $g_c\neq0$. For $g_c=0$, the CPW resonator is decoupled from the flux qubit and Eq.~(\ref{spectrum2}) is then simplified to $S_V^0(\omega)=\frac{\kappa}{(\kappa/2)^2+(\omega-\omega_c)^2}$, i.e., a Lorentzian shape with a peak at $\omega=\omega_c$~[see the solid curve in Fig.~2(a)]. For $g_m=0$ but $g_c\neq0$, the VFS in Eq.~(\ref{spectrum2}) can also be simplified to a form similar to $S_V^0(\omega)$, but the peak has red-shifted to $\omega=\omega_c^\prime\equiv\omega_c-g_c^2/\Delta_c$ due to the flux qubit~[see the dashed curve in Fig.~2(a)]. This frequency shift can be understood as a result of the dispersive interaction between the flux qubit and the CPW resonator, which induces a qubit-state-dependent frequency shift in the CPW resonator or a photon-number-dependent frequency shift in the flux-qubit transition frequency. As we concentrate on the CPW resonator and trace out the degrees of freedom of the flux qubit initially in the ground state, we can observe the red-shift of the frequency for the CPW resonator.

When the coupling strength $g$ is nonzero, a vacuum Rabi splitting can be observed from the VFS affected by the NAMR. From Fig.~2(b), we can explicitly find that the single peak in Fig.~2(a) is split to two resolved peaks located at $\omega_{\pm}=\omega_{0}-(\frac{g_c^2+g_m^2}{2}\mp\sqrt{g_c^4-g_c^2g_m^2+g_m^4})/\Delta_{0}$ in the resonance case of $\omega_c=\omega_m=\omega_0$ or equivalently $\Delta_a=\Delta_c=\Delta_0$. These two peaks are separated by $\delta \omega\equiv\omega_+-\omega_-=2\sqrt{g_c^4-g_c^2g_m^2+g_m^4}/\Delta_{0}$ rather than two times the coupling strength in the two-body system such as an atom-cavity system~\cite{Thompson}. The underlying physics is that the peaks of the CPW resonator and the NAMR are first shifted by the flux qubit due to the dispersive interaction, and then the states of the CPW resonator are dressed by the NAMR because of the induced effective coupling $g$. That is why two peaks can be observed in Fig.~2(b). Note that the reduced NAMR-CPW resonator system governed by the Hamiltonian~(\ref{Reduced H}) is not in resonance when $\omega_c=\omega_m$, since the modified frequencies of the NAMR and the CPW resonator are associated with the coupling strengths $g_c$ and $g_m$. In our proposed hybrid system, while the coupling $g_c$ between the CPW resonator and the flux qubit is fixed, the coupling $g_m$ is however tunable via the externally applied magnetic field $B$. In Fig.~2(b), we also show the VFS of the CPW resonator for two different values of the coupling strength $g_m$. These results reveal that as $g_m$ increases, the height of the left peak increases but that of the right peak decreases. This phenomena is related to the asymmetry of the coupling strengths $g_m$ and $g_c$.

\begin{figure}
    \includegraphics[scale=0.55]{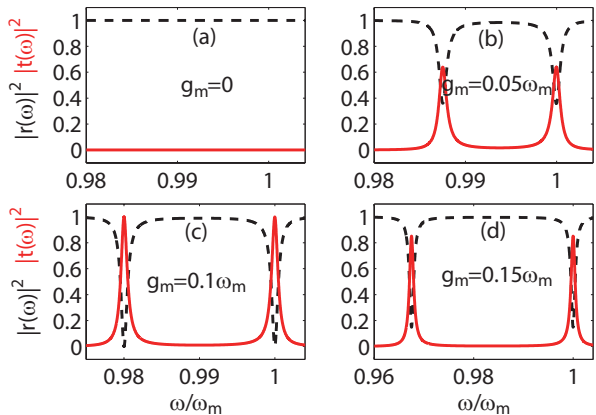}\\
  \caption{(Color online) The reflectance $|r(\omega)|^2$ and the transmittance $|t(\omega)|^2$ as a function of the normalized frequency $\omega/\omega_m$ for various values of coupling strength $g_m$ and $\omega_c/2\pi=\omega_m/2\pi=1$GHz. Other parameters are the same as in Fig.~2.}\label{f3}
\end{figure}

\begin{figure}
    \includegraphics[scale=0.55]{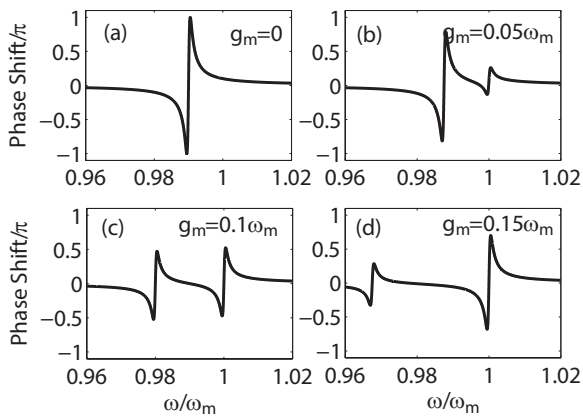}\\
  \caption{(Color online) The phase shift as a function of the normalized frequency $\omega/\omega_m$ for various values of coupling strength $g_m$. Other parameters are the same as in Fig.~3. }
\end{figure}


\section{Single photon transport}

Below we further study the properties of a single photon transport in this hybrid system using the standard input-output theory~\cite{Walls}. With Eq.~(\ref{operator}) available, we can directly write the reflection coefficient as
\begin{equation}
r(\omega)=\frac{c_{\rm out}}{c_{\rm in}}=\frac{\left[ \kappa/2
+i(\omega -\omega_{c}^\prime)\right] \left[ \gamma/2-i(\omega -\omega_{m}^\prime)\right]
-g^{2}}{\left[ \kappa/2-i(\omega -\omega_{c}^\prime)\right] \left[ \gamma/2-i(\omega -\omega_m^\prime)\right] +g^{2}},\label{reflection}
\end{equation}
where we have assumed that the input field $b_{\rm in}$ of the NAMR at a low temperature provides negligible contributions to the output field $c_{\rm out}$ of the CPW resonator. Note that Eq.~(\ref{reflection}) is a general expression for arbitrary values of $g$. In the case of a vanishing $g$, the reflection coefficient in Eq.~(\ref{reflection}) reduces to
\begin{equation}
r_{0}(\omega )=\frac{\kappa/2+i(\omega -\omega_{c}^\prime)}{\kappa/2-i(\omega -\omega_{c}^\prime)},
\end{equation}%
The result indicates that the flux qubit significantly detuned with the CPW resonator dose not affect the reflectance. Thus the incident photon has a perfect reflection in the whole frequency range because the system is equivalent to a bare resonator [see the black dashed line in Fig.~3(a)].

In the case of $g\neq0$, the reflection coefficient has two dips due to the effective coupling of the NAMR to the CPW resonator. In Figs.~3(b)$-$3(d), we plot the transmission and reflection coefficients as a function of the normalized frequency $\omega/\omega_m$ for three values of $g_m$, where transmission coefficient is defined as $|t(\omega)|^2=1-|r(\omega)|^2$. From Figs.~3(b) and 3(d), we can find that the incident photon propagating in the CPW resonator will have partial transmission and reflection due to the asymmetric coupling strengths of $g_c$ and $g_m$ leading to $\omega_c^\prime\neq\omega_m^\prime$ for the reduced NAMR-CPW resonator system described by Eq.~(\ref{Reduced H}). In Fig.~3(c), symmetric coupling strengths, i.e., $g_c=g_m$, are considered. The effective frequencies for the NAMR and the CPW resonator are in resonance, i.e., $\omega_c^\prime=\omega_m^\prime$. In this case, the photon transmission coefficient can go up to one with vanishing reflection at $\omega=\omega_c-g_c^2/\Delta_c\pm\sqrt{(g_cg_m/\Delta_c)^2-\kappa\gamma/4}$, corresponding to the right dip for `$+$' and the left dip for `$-$'. That is, the transmission and reflection coefficients can be tuned from zero to one by tuning the coupling strength $g_m$ via the external magnetic field $B$, while the coupling strength $g_c$ is  fixed. The phenomenon implies that an incident single photon propagating in the CPW resonator has two possible output ports depending on how we tune the external magnetic field. Therefore, a quantum router can be realized. As we know, a single photon propagating in the CPW resonator will acquire a phase shift corresponding to the imaginary part of the reflection coefficient in Eq.~(\ref{reflection}). We show that a single photon traveling in the CPW resonator containing a flux qubit at a large detuning will acquire a phase shift of $\pm\pi$ at $\omega=\omega_c-g_c^2/\Delta_c$. However, this phase shift will reduce to zero rapidly as the frequency of the incident photon deviated from the resonant point [see Fig. 4(a)]. In the presence of the NAMR, we find that this spectral feature in the phase shift splits into two and their positions can be tuned by controlling the external magnetic field. In Fig.~4(b)$-$4(d), we plot the acquired phase shift as a function of the normalized frequency at a finite coupling strength $g_m$. When the coupling strength $g_m$ is increased, the new feature on the right-hand side increases in amplitude. Inversely, the one on the left-hand side becomes less pronounced. Investigation of the acquired phase shift can be employed to further study group delay $\tau_d$ of the photon given by
$\tau_d=\partial \phi/\partial\omega$ \cite{MTCheng}, where $\phi$ is defined by rewriting the reflection coefficient as $r(\omega)=\sqrt{|r(\omega)|^2}\text{exp}(i\phi)$.

\section{The effect of the qubit decay}

\begin{figure}[H]
  \includegraphics[scale=0.55]{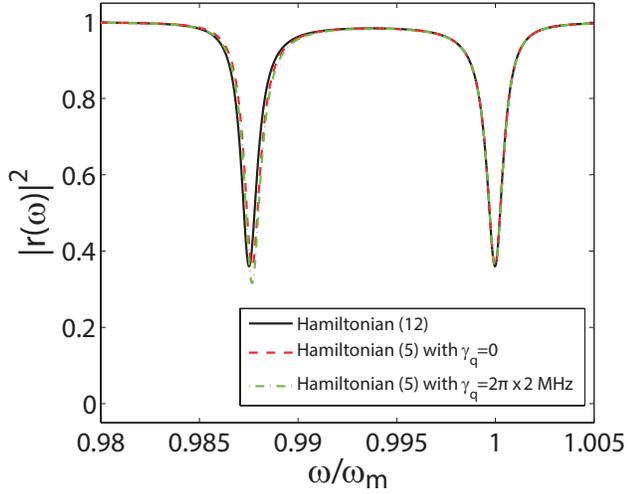}\\
  \caption{(Color online) The reflection coefficient as a function of the normalized frequency $\omega/\omega_m $, with $\langle\sigma_z\rangle=-1$. The solid black curve corresponds to the reflection coefficient obtained using the effective Hamiltonian (\ref{Reduced H}), and the other two curves (the dashed red curve and the dot-dashed green one) correspond to the reflection coefficient obtained using the original Hamiltonian (\ref{total H}). The parameters are the same as in Fig.~\ref{f3}(b).}\label{f5}
\end{figure}

In the above study, we have ignored the effect of the qubit decay induced by the environment of the qubit. Below we show this approximation is reasonable. Here we start with the original Hamiltonian (\ref{total H}) to calculate the reflection coefficient $r(\omega)$ of the incident photon by including the qubit decay $\gamma_q$. The quantum Langevin equations in Eq.~(\ref{QLE}) are modified as

\begin{align}
\frac{dc}{dt}=&-(i\omega_c+\frac{\kappa}{2}) c-ig_c \sigma_-+\sqrt{\kappa}c_{\rm in},\nonumber\\
\frac{db}{dt}=&-(i\omega_m+\frac{\gamma}{2}) b-ig_m \sigma_-+\sqrt{\gamma}b_{\rm in},\\
\frac{d\sigma_-}{dt}=&-(i\omega_q+\frac{\gamma_q}{2}) \sigma_-+i(g_c c+g_m b) \sigma_z.\nonumber
\end{align}

\begin{figure}
  \includegraphics[scale=0.55]{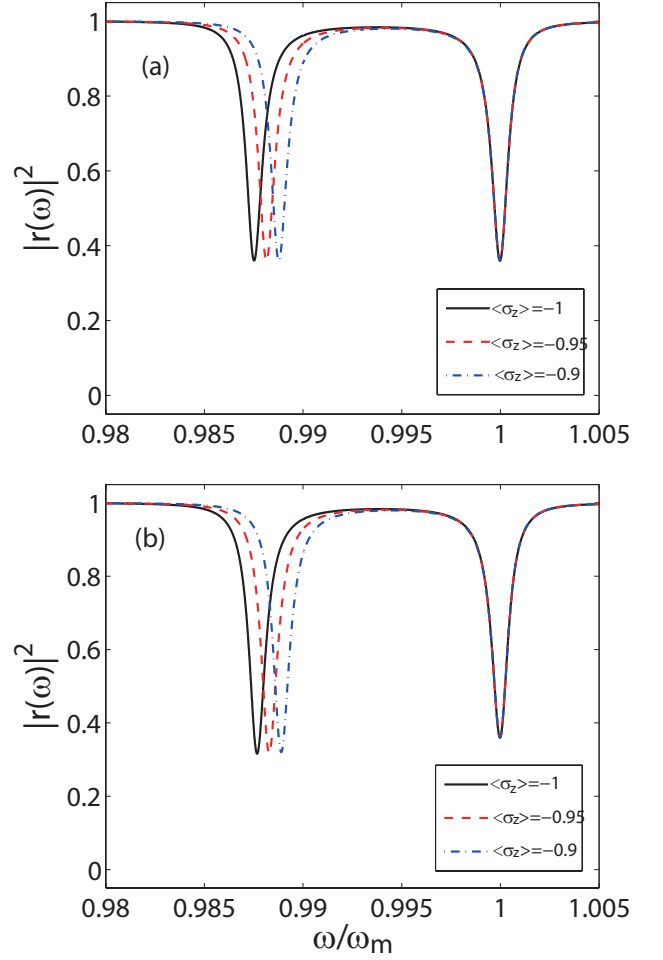}\\
  \caption{(Color online) The reflection coefficient as a function of the normalized frequency $\omega/\omega_m$ by considering the thermal fluctuation. (a) The results obtained using the effective Hamiltonian (\ref{Reduced H}) when the qubit decay is not included; (b) The results obtained using the original Hamiltonian (\ref{total H}) when the qubit decay is included. In (b) we choose $\gamma_q=2\pi\times2$ MHz, and other parameters are the same as in Fig.~\ref{f3}(b).}\label{f6}
\end{figure}
Then, the operator $c(\omega)$ in Eq.~(\ref{operator}) becomes

\begin{align}
c(\omega)=\frac{A_m A_q\sqrt{\kappa} c_{\rm in}-g_m^2\sqrt{\kappa}\sigma_z c_{\rm in}+g_c g_m \sigma_z \sqrt{\gamma}b_{\rm in}}{A_c A_m A_q
-A_m g_c^2 \sigma_z
-A_c g_m^2 \sigma_z
},\label{C}
\end{align}
where $A_c=\frac{\kappa}{2}-i(\omega-\omega_c)$, $A_m=\frac{\gamma}{2}-i(\omega-\omega_m)$, and $A_q=\frac{\gamma_q}{2}-i(\omega-\omega_q)$. By tracing over the degrees of freedom of the qubit, the reflection coefficient $r(\omega)$ can be written as

\begin{align}
r(\omega)=&\frac{c_{\rm out}}{c_{\rm in}}=\frac{(A_m A_q-g_m^2\langle\sigma_z\rangle)A_c^*+g_c^2 A_m\langle\sigma_z\rangle}{(A_m A_q-g_m^2\langle\sigma_z\rangle)A_c-g_c^2 A_m\langle\sigma_z\rangle},\label{R'}
\end{align}
where $A_c^*$ is the conjugate of $A_c$. As assumed in Sec. III, the flux qubit is initially prepared in the ground state. Because the qubit is in the dispersive regime with the NAMR and the CPW resonator, it is kept in the ground state for a low temperature (e.g., $T\sim20 $ mK), thus, $\langle\sigma_z\rangle=-1$ in Eq.~(\ref{R'}).

In Fig.~\ref{f5}, we plot the reflection coefficient as a function of the normalized frequency $\omega/\omega_m$, where $\langle\sigma_z\rangle=-1$. The solid curve corresponds to the reflection coefficient obtained using the effective Hamiltonian (\ref{Reduced H}) and the other two curves correspond to the reflection coefficient obtained using the original Hamiltonian (\ref{total H}), with $\gamma_q=0$ (dashed curve) and $2\pi\times2$ MHz (dot-dashed curve). We can see that there exists a small frequency shift between the solid and dashed curves. This difference arises from the second-order approximation employed for deriving the effective Hamiltonian (\ref{Reduced H}). However, while this difference is appreciable for small values of frequency, it is greatly reduced in the large frequency regime (see Fig.~\ref{f5}). Furthermore, we study the effect of the thermal fluctuation on the reflection coefficient. For the reflection coefficient obtained using the effective Hamiltonian (\ref{Reduced H}), the thermal fluctuation can only result in a small frequency shift to the left peak of the reflection coefficient and the frequency shift to the right peak is negligible [see Fig.~\ref{f6}(a)]. Also, the situation is similar for the reflection coefficient obtained using the original Hamiltonian (\ref{total H}) [see Fig.~\ref{f6}(b)]. These numerical results indicate that the approximation employed in deriving the effective Hamiltonian (\ref{Reduced H}) is reasonable.

\section{Conclusion}

In conclusion, we have proposed a hybrid system which can achieve a strong and tunable coupling between a NAMR and a CPW resonator mediated via a  flux qubit. With this setup, the transfer of quantum information between the NAMR and the CPW resonator can be accomplished. We have also deduced a  vacuum Rabi splitting of a resonant peak in the voltage-fluctuation spectrum of the CPW resonator. It also leads to a split in the spectral features in the transmittance and the phase shift spectrum. By tuning the external magnetic field, the transmittance can be tuned from 0 to 1, indicating the possible application as a quantum router.

This work is supported by the NSAF No.~U1330201, the NSFC No.~91421102, the NBPRC No.~2014CB921401, Hong Kong GRF~(Grant No.~501213). 

\end{document}